\documentclass[argument]{aastex61}
\usepackage{multirow}
\usepackage{amssymb,amsfonts,amsmath}

\usepackage{graphicx}
\usepackage{epstopdf}
\usepackage{lineno}
\usepackage{graphicx}
\usepackage{textgreek}
 \received{-------}
\accepted{-------}

\shortauthors{Abdollahi et al.}
\begin{document}

\title{On the origin of the gamma-ray emission toward SNR CTB 37A with \textit{Fermi}-LAT}

\correspondingauthor{Soheila Abdollahi}
\email{abdollahi.soheila@gmail.com}

\author[0000-0002-0786-7307]{Soheila Abdollahi}

\affil{Department of Physical Sciences, Hiroshima University, Higashi-Hiroshima, Hiroshima 739-8526, Japan}

\author{Jean Ballet}
\affiliation{AIM, CEA, CNRS, Universit$\acute{e}$ Paris-Saclay, Universit$\acute{e}$ Paris Diderot, Sorbonne Paris Cit$\acute{e}$, F-91191 Gif-sur-Yvette, France}

\author{Yasushi Fukazawa}
\affiliation{Department of Physical Sciences, Hiroshima University, Higashi-Hiroshima, Hiroshima 739-8526, Japan}

\author{Hideaki Katagiri}
\affiliation{College of Science, Ibaraki University, 2-1-1, Bunkyo, Mito 310-8512, Japan}

\author{Benjamin Condon}
\affiliation{Centre d'$\acute{E}$tudes Nucl$\acute{e}$aires de Bordeaux Gradignan, IN2P3/CNRS, Universit$\acute{e}$ Bordeaux 1, BP120, F-33175 Gradignan Cedex, France}

\begin{abstract}
The middle-aged supernova remnant (SNR) CTB 37A is known to interact with several dense molecular clouds through the detection of shocked ${\rm H_{2}}$ and OH 1720 MHz maser emission. In the present work, we use eight years of $\textit Fermi$-LAT Pass 8 data, with an improved point-spread function and an increased acceptance, to perform detailed morphological and spectral studies of the $\gamma$-ray emission toward CTB 37A from 200 MeV to 200 GeV. The best fit of the source extension is obtained for a very compact Gaussian model with a significance of 5.75\,$\sigma$ and a 68\% containment radius of $0\fdg116$\,$\pm$\,$0\fdg014_{\rm stat}$\,$\pm$\,$0\fdg017_{\rm sys}$ above 1 GeV, which is larger than the TeV emission size. The energy spectrum is modeled as a LogParabola, resulting in a spectral index $\alpha$ = 1.92\,$\pm$\,0.19 at 1 GeV and a curvature $\beta$ = 0.18\,$\pm$\,0.05, which becomes softer than the TeV spectrum above 10 GeV. The SNR properties, including a dynamical age of 6000 yr, are derived assuming the Sedov phase. From the multiwavelength modeling of emission toward the remnant, we conclude that the nonthermal radio and GeV emission is mostly due to the reacceleration of preexisting cosmic rays (CRs) by radiative shocks in the adjacent clouds. Furthermore, the observational data allow us to constrain the total kinetic energy transferred to the trapped CRs in the clouds. Based on these facts, we infer a composite nature for CTB 37A to explain the broadband spectrum and to elucidate the nature of the observed $\gamma$-ray emission. 
\end{abstract}
\keywords{acceleration of particles --- (ISM:) cosmic rays --- ISM: supernova remnants --- radiation mechanisms: non-thermal}
%%%%%%%%%%%%%%%
\section{Introduction} \label{sec:intro}
Supernova remnants (SNRs) have been regarded as the most promising candidates for the bulk of Galactic cosmic rays (CRs).~Energetic particles are produced at the shock waves associated with SNRs through the diffusive shock acceleration process~\citep[DSA;][]{Axford.Leer.Skadron77, Krymskii77, Bell78a, Bell78b, Blandford.Ostriker78}. The main $\it phenomenological$ argument in favor of this hypothesis is that SNRs are able to provide the total energy budget necessary to maintain the Galactic population of CRs, if approximately 10\% of the kinetic energy released by supernova (SN) explosions can be transferred to CRs at SNR shocks~\citep{Ginzburg.Syrovatskii64}. Moreover, the DSA mechanism can explain the hard power-law energy spectrum of CRs at their source with differential spectral index close to 2~\citep[e.g.,][]{Fermi49, Bell78a, Bell87}. \par
The recent observations of several (young and middle-aged) SNRs in very high-energy (VHE) $\gamma$-rays ($\geq$ 100 GeV) as well as nonthermal X-ray emission imply effective production of relativistic particles in support of the DSA paradigm, although the radiation mechanism responsible for the GeV/TeV emission is still under debate. This is because $\gamma$-rays can be produced by energetic hadrons (protons and nuclei) through inelastic collisions with interstellar gas, and/or by energetic electrons through nonthermal bremsstrahlung and inverse Compton (IC) scattering of ambient radiation fields. Therefore, to identify the nature of the $\gamma$-ray emission and particle species, detailed morphological and spectral studies correlated with other multiwavelength data are crucial. \par 
Middle-aged SNRs interacting with nearby dense molecular clouds (MCs) show strong evidence of relativistic protons. The characteristic neutral pion-decay signature in the $\gamma$-ray spectrum below $\sim$200 MeV (often called \textquotedblleft pion~bump\textquotedblright) in three SNRs, IC 443, W44, and W51, associated with MCs is explicitly linked to the hadronic acceleration~\citep{Ackermann13, Jogler.Funk16}. Additional support for a hadronic origin of the $\gamma$-ray emission from SNRs interacting with MCs (often termed SNR/MC interactions) comes from observations of W28 \citep{Abdo10}, W41, MSH 17$-$39, G337.7$-$0.1 \citep{Castro13}, and G5.7$-$0.1 \citep{Joubert16}. The SNR$\slash$MC interactions have been confirmed through the detection of 1720 MHz OH maser emission toward shocked regions of these remnants \citep[e.g.,][]{Frail96}. Given the ambient gas density in the middle-aged SNRs, the total energy budget in accelerated particles is estimated to be $\simeq 10^{50}$ erg. \par 
SNR CTB 37A (also known as G348.5+0.1) was initially identified as a discrete source in radio surveys by~\citet{Milne.Hill69}. It is a middle-aged remnant, $\sim$$10^{4}$ yr old~\citep{Sezer11, Pannuti14}, located in the CTB 37 complex region near two other remnants, CTB 37B (G348.7+0.3, associated with HESS J1713$-$381) and G348.5$-$0.0. The distance to the remnant has been estimated to be in the range between 6.3 and 9.5 kpc, based on H\,{\footnotesize I} absorption measurement along the line of sight in high-resolution Southern Galactic Plane Survey data by~\citet{Tian.Leahy12}. The distances obtained for CTB 37B and G348.5$-$0.0 ($\sim$13.2 kpc and $\leq$~6.3 kpc, respectively) by those authors imply that the three remnants of the CTB 37 complex merely lie along similar lines of sight while they differ in distance. Association of CTB 37A with several nearby dense MCs has been firmly established based on observations of several OH (1720 MHz) maser spots detected toward the remnant~\citep{Frail96} and also shocked clumps of clouds with high column densities~\citep{Reynoso.Mangum00, Maxted13, Braiding18}. \par 
CTB 37A has been observed extensively across a wide range of energies, from radio to VHE $\gamma$-rays. Radio observations of the SNR~\citep{Milne.Dickel75, Dawnes84, Kassim91} have revealed a shell structure with bright emission in the northern and eastern rims, and also a faint extension suggestive of a \textquotedblleft breakout\textquotedblright~into an inhomogeneous medium with a large-scale density gradient toward the southwest~\citep{Kassim91}.~The angular size of the remnant (including the breakout) is estimated to be 19$\times$16 arcmin$^{2}$ at $\nu$ = 843 MHz by~\citet{Whiteoak.Green.96}. From the same radio observations, a flux density of $S_{\nu}$ = 71 Jy is obtained by those authors. \citet{Kassim91} derived a typical spectral index of $\alpha_{\rm r}$ = $-$0.5~$\pm$~0.1 (where $\it S_{\nu} \propto \nu^{\alpha_{\rm r}}$) at frequencies above 330 MHz.~In the X-ray band, in addition to the extended soft thermal component dominated by emission lines of highly ionized species of Mg, Si, S, Ar, and Ca, a compact nonthermal hard X-ray emission has also been detected from the northwest of the SNR, which might be associated with the emission of a pulsar wind nebula (PWN) as suggested by~\citet{Aharonian08},~\citet{Sezer11}, and~\citet{Yamauchi14}.~Searches for $\gamma$-ray pulsations in this region by~\citet{SazParkinson18} led to the discovery of pulsar PSR J1714$-$3830 coincident with~SNR CTB~37A.~X-ray and radio observations of this source indicate that CTB 37A is a mixed-morphology SNR~\citep{Aharonian08} as characterized by a center-filled thermal X-ray emission surrounded by a shell-like radio structure.~At GeV energies, it was detected for the first time as an extended source of Gaussian width $0\fdg13$ with a significance of $\sim$4.5\,$\sigma$ by~\citet{Brandt13}.~In the previous study, by~\citet{Castro.Slane10}, a detailed characterization of the spectral properties was not possible due to limited photon statistics.~The TeV $\gamma$-ray source HESS J1714$-$385 with extension of $\sim$0\fdg07 is positionally coincident with the SNR~\citep{Aharonian08}, though the hadronic or leptonic nature of the $\gamma$-ray emission toward this source still remains elusive. \par
In this paper,~benefiting from a significant improvement in the LAT sensitivity implemented in Pass 8 event selection/reconstruction as well as increased photon statistics, we report a detailed analysis of 8 yr of \textit{Fermi}-LAT $\gamma$-ray data around the SNR CTB 37A, and discuss the morphological and spectral characteristics of the $\gamma$-ray emission toward the remnant, which are crucial for distinguishing between hadronic and leptonic scenarios.~In Section~\ref{sec:ObsDataRed}, observations and data reduction are briefly described. The analysis procedures and results are given in Section~\ref{sec:AnalysisResults}, where the spatial extension and spectrum of the remnant are explained. In Section~\ref{sec:Disc}, we argue that the SNR is a composite system. In addition, the crushed clouds scenario is examined to explain the multiwavelength emission toward the system through the reacceleration of trapped CRs by the radiative shocks and the subsequent adiabatic compression. A comparison to other SNRs is presented in Section~\ref{sec:Comparison}. Finally, the conclusions are summarized in Section~\ref{sec:Conclusions}.
%%%%%%%%%%%%%%%%%%%%%%%%%%
\section{Observations and Data Reduction}\label{sec:ObsDataRed}
The \textit{Fermi} Large Area Telescope (LAT) is a pair-conversion detector, designed to survey the high-energy $\gamma$-ray sky in the energy range from $\sim$20 MeV to above 300 GeV. The LAT is equipped with a tracker$\slash$converter for direction reconstruction of incident $\gamma$-rays, a CsI(Tl) crystal calorimeter for measurement of the energy deposition, and a surrounding anticoincidence detector for rejection of the charged particle background. Full details of the LAT instrument and data processing can be found in~\citet{Atwood09}, and information regarding the on-orbit calibration is given in~\citet{Abdo09a}. \par 
Recently, new event reconstruction and classification known as Pass 8~\citep{Atwood13}, released by the \textit{Fermi}-LAT collaboration, has provided substantial improvements in the instrument response functions (IRFs). Pass 8 presents significantly increased effective area ($\sim$8000 cm$^{2}$ on-axis above 1 GeV), an improved angular resolution as given by the point-spread function (PSF, with a 68\% containment radius of $\sim$0$\fdg$8 at 1 GeV), a reduced energy dispersion ($<$10\% between 1 GeV and 100 GeV), and a wider field of view ($\sim$2.4 sr at 1 GeV). Taking advantage of the Pass 8 data combined with the increased photon statistics collected by the LAT, more information on source extension and spectral properties will be revealed. \par
The LAT Pass 8 data used for the following analysis were collected in sky-survey mode during the first eight years of scientific operations, which began on 2008 August 4. The $\gamma$-rays in the 0.2$-$200~GeV energy range within a region of interest (ROI) with a radius of 15$^{\circ}$ centered on the position of CTB 37A are selected for the binned maximum likelihood analysis.~The 200 MeV lower limit was chosen to reduce the contamination from underpredicted $\gamma$-ray emission at low energies in the Galactic plane.~The event selection is based on the low background Pass 8 $\texttt{source}$ event class and the corresponding IRF is $\texttt{P8R2\char`_SOURCE\char`_V6}$.~A zenith angle cut of 90$^{\circ}$ and 105$^{\circ}$ for events below and above 1 GeV, respectively, is applied to minimize the contamination from the Earth limb~\citep{Abdo09b}. 
%%%%%%%%%%%%%%%%%%%%%%%%%%
\section{Analysis and Results} \label{sec:AnalysisResults}
The $\gamma$-ray data were analyzed using the LAT $\texttt{Science~Tools}$ software package (v11r05p02), publicly available from the Fermi Science Support Center (FSSC).\footnote{The Science Tools package and supporting documents are distributed by the Fermi Science Support Center and can be accessed at \url{http://fermi.gsfc.nasa.gov/ssc/data/analysis/software/}.} Two different tools ($\texttt{pointlike}$ and $\texttt{gtlike}$) were used to determine the morphological and spectral characteristics of the remnant, respectively, under the maximum likelihood fitting technique~\citep{Mattox96}.~$\texttt{pointlike}$ is an alternative package for fast binned likelihood analysis~\citep{Kerr10} validated by~\citet{Lande12} and is optimized to evaluate the best-fit position and extension of the source before performing a more accurate fit of the spectrum using $\texttt{gtlike}$. These tools fit a $\gamma$-ray emission model to the LAT data along with models for the instrumental, extragalactic, and Galactic components of the background. The Galactic diffuse emission is modeled using the LAT standard diffuse emission model \texttt{gll\char`_iem\char`_v06.fits}, and the residual instrumental background and extragalactic $\gamma$-ray radiation are combined into a single isotropic component with a spectral shape described by a tabulated model \texttt{iso\char`_P8R2\char`_SOURCE\char`_V6\char`_v06.txt}. The models and their detailed descriptions are available from the FSSC. In addition to the two aforementioned diffuse models and all background sources within 20$^{\circ}$ around CTB 37A listed in the third \textit{Fermi}-LAT catalog~\citep[3FGL,][]{Acero2015}, two statistically significant point sources at a distance $<$ 0$\fdg$5 from CTB 37A, not already detected by the 3FGL, are included in the region model. One of them, the CTB 37B source (3FHL J1714.0$-$3811) with a power-law spectrum, detected by the LAT at energies above 10 GeV, is taken from the third catalog of hard \textit{Fermi}-LAT sources~\citep[3FHL,][]{Ajello17}. The other one, the FL8Y J1714.8$-$3850 source with a LogParabola spectral model, located in the southeast of the remnant, is given by the preliminary \textit{Fermi}-LAT eight-year list of sources (FL8Y)\footnote{\url{https://fermi.gsfc.nasa.gov/ssc/data/access/lat/fl8y/}}.~The detected pulsar PSR J1714-3830 is not included in the model because the ephemerides are not long enough to cover the entire LAT data set. Therefore, it is assumed that the entire $\gamma$-ray emission is originated in the SNR, which establishes an upper limit on the GeV flux of the remnant. Detailed analyses of off-pulse emission will be discussed in future work. Background sources over an area 5$^{\circ}$ larger than the ROI are contained in the model to account for the contamination from their emission. \par
As a first step, to assess the morphology of CTB 37A, the position and extension of the remnant, together with the position of the two nearby point sources, CTB 37B and FL8Y J1714.8$-$3850, were fitted in an iterative process. Moreover, in this approach, a preliminary estimate of the spectral values for sources within 2$^{\circ}$ of CTB 37A, which may affect its flux, was provided by allowing their spectral parameters (normalization, index, and cutoff energy) to vary. Only for the source 3FGL J1718.1$-$3825, located $0\fdg7$ away from the remnant and associated with the pulsar PSR J1718$-$3825, our checks showed that the pulsar is not bright enough to justify freeing the exponential index (\textit{b}) of an exponentially cutoff power-law model, and so, \textit{b} was fixed to 1. All remaining spatial and spectral parameters of sources in the initial ROI model were kept fixed at the 3FGL values. Once the spatial characteristics of the SNR had been approximated, as a second step, a global fit by $\texttt{gtlike}$ was performed using the best-fit spatial model of the first step to determine a slightly more accurate estimate of the spectral parameters previously fitted by $\texttt{pointlike}$. Then, as a final step, to measure the energy spectrum of CTB 37A, the $\gamma$-ray data were fitted to the model in narrow energy bins. All parameters except the normalization of the bright sources within 1$\fdg$3 from the ROI center (including CTB 37A, CTB 37B, RX J1713.7$-$3946, FL8Y J1714.8$-$3850, 3FGL J1718.1$-$3825, and 3FGL J1718.0$-$3726) were fixed to those previously found in the global fit. Doing so helps with avoiding numerical instabilities resulting from the fine binning in energy. During all the analyses, the normalizations of both the Galactic and isotropic diffuse emissions were left free to account for uncertainties in the diffuse emission. \par 
For the following analyses, three primary sources of systematic errors have been taken into account: (1) uncertainties due to imperfect modeling of the Galactic diffuse emission, $\sigma_{\rm IEM}$; (2) uncertainties on the source spatial model, $\sigma_{\rm model}$; and (3) uncertainties on the PSF, $\sigma_{\rm PSF}$. Systematic errors in the Galactic diffuse emission model were evaluated by going over the whole process using an alternative diffuse emission model~\citep[the Sample model used in][]{Ackermann17}. The systematic errors associated with spatial modeling of the source were obtained by comparing the SNR's estimated properties for the disk and Gaussian models provided by $\texttt{pointlike}$. Finally, for the $\texttt{P8R2\char`_SOURCE\char`_V6}$ event class in our analysis, systematics on the PSF 68\% containment radius were estimated to be $<$ 5\% between 100 MeV and 10 GeV, increasing to 25\% at 1 TeV as explained in the caveat page of the FSSC\footnote{\url{https://fermi.gsfc.nasa.gov/ssc/data/analysis/LAT_caveats.html}}, which results in a systematic uncertainty of $0\fdg005$ on the 68\% containment radius following~\citet{Ackermann18}. Further details on the analyses and results are given in the subsequent subsections. 
%%%%%%%%%%%%%%%%%%%%%%%%%%
\subsection{Morphological Analysis} \label{subsec:Morphology}
The spatial analysis of the source was performed using all events above 1~GeV, taking advantage of a narrower instrumental PSF in conjunction with the significantly reduced contamination by the Galactic diffuse emission. The position and possible spatial extension of CTB 37A were examined using $\texttt{pointlike}$ assuming a LogParabola spectral shape, $dN/dE=N_{0}(E/E_{0})^{-(\alpha\,+\,\beta\, \rm{ln}(\it{E/E}_{\rm 0}))}$, as described in the 3FGL. Starting with the point-source hypothesis as our baseline model, we individually fitted the position and spectrum of CTB 37A and the two closest sources, CTB 37B and FL8Y J1714.8$-$3850. The iterative fitting process was progressively continued until changes in the estimated parameters from two consecutive fits became sufficiently small, and the derivative of the log-likelihood was close to zero (here, delta log-likelihood $<$ 0.1).~To examine the possible angular extension of the SNR, we then replaced the point source on CTB 37A with an extended source (using a uniform disk and a 2D symmetric Gaussian model) and repeated the procedure until the desired fit with stable estimates was achieved.~The significance of the source extension is quantified as the test statistics TS$_{\rm ext}$ = 2~$\mathrm{log}$($\mathcal{L}$$_{\mathrm{ext}}$/$\mathcal{L}$$_{\mathrm{ps}}$), which compares the likelihood of an extended source hypothesis with a point-source one. For one additional degree of freedom, the extension significance is $\sqrt{\rm TS}$ in units of $\sigma$. \par
\begin{figure*}[!ht]
\gridline{\fig{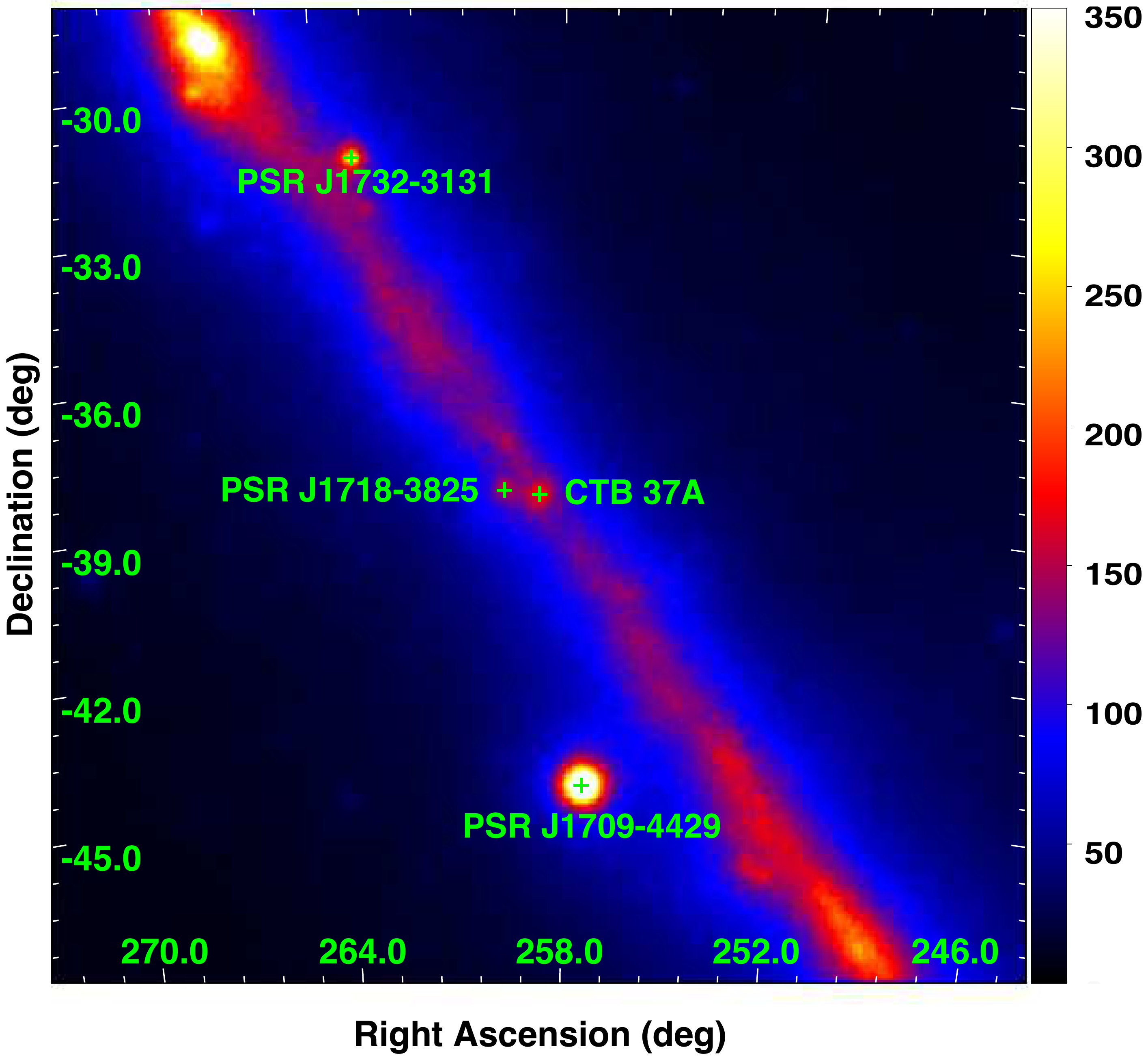}{0.40\textwidth}{}
          \fig{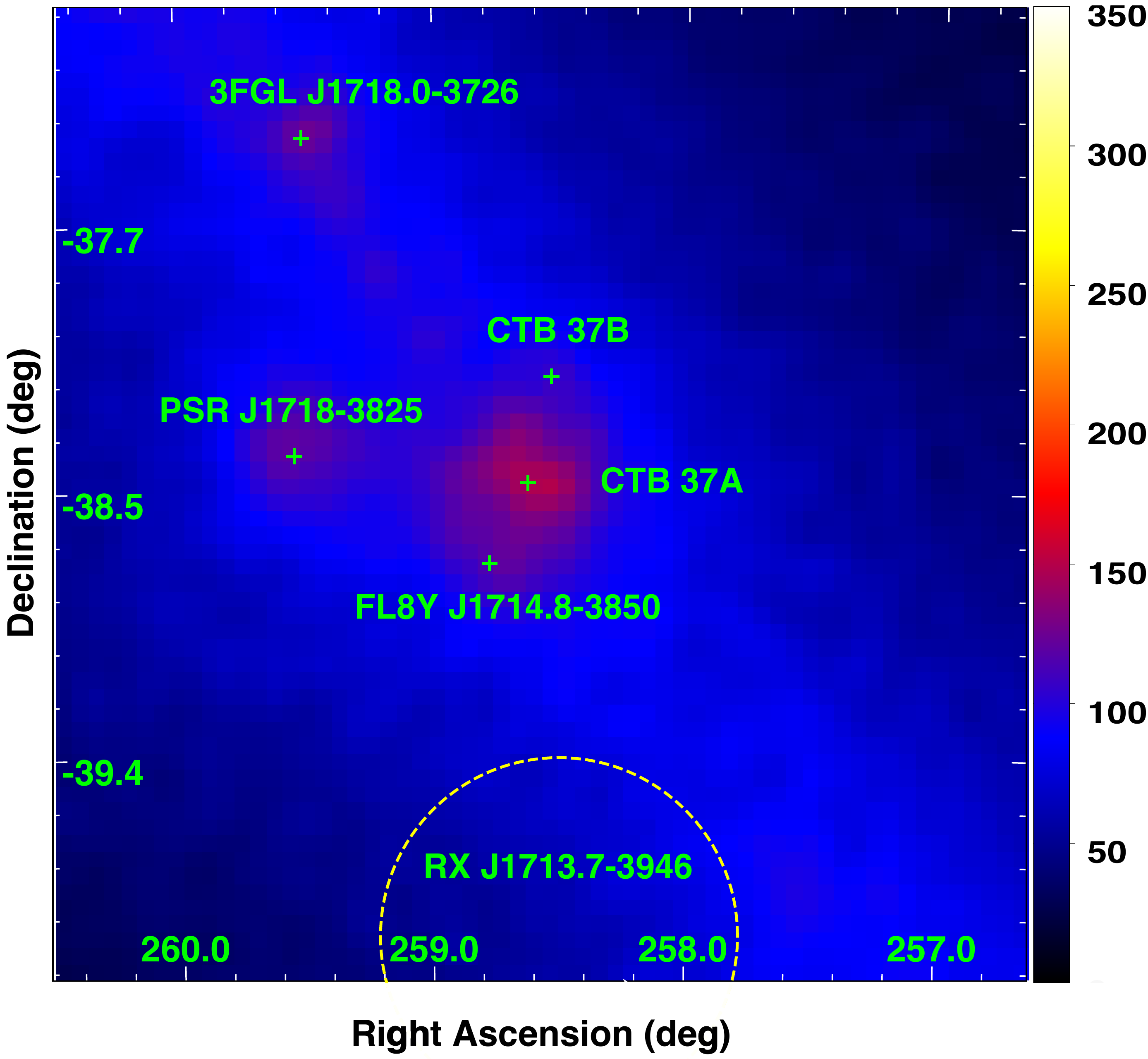}{0.40\textwidth}{}}
\gridline{ \fig{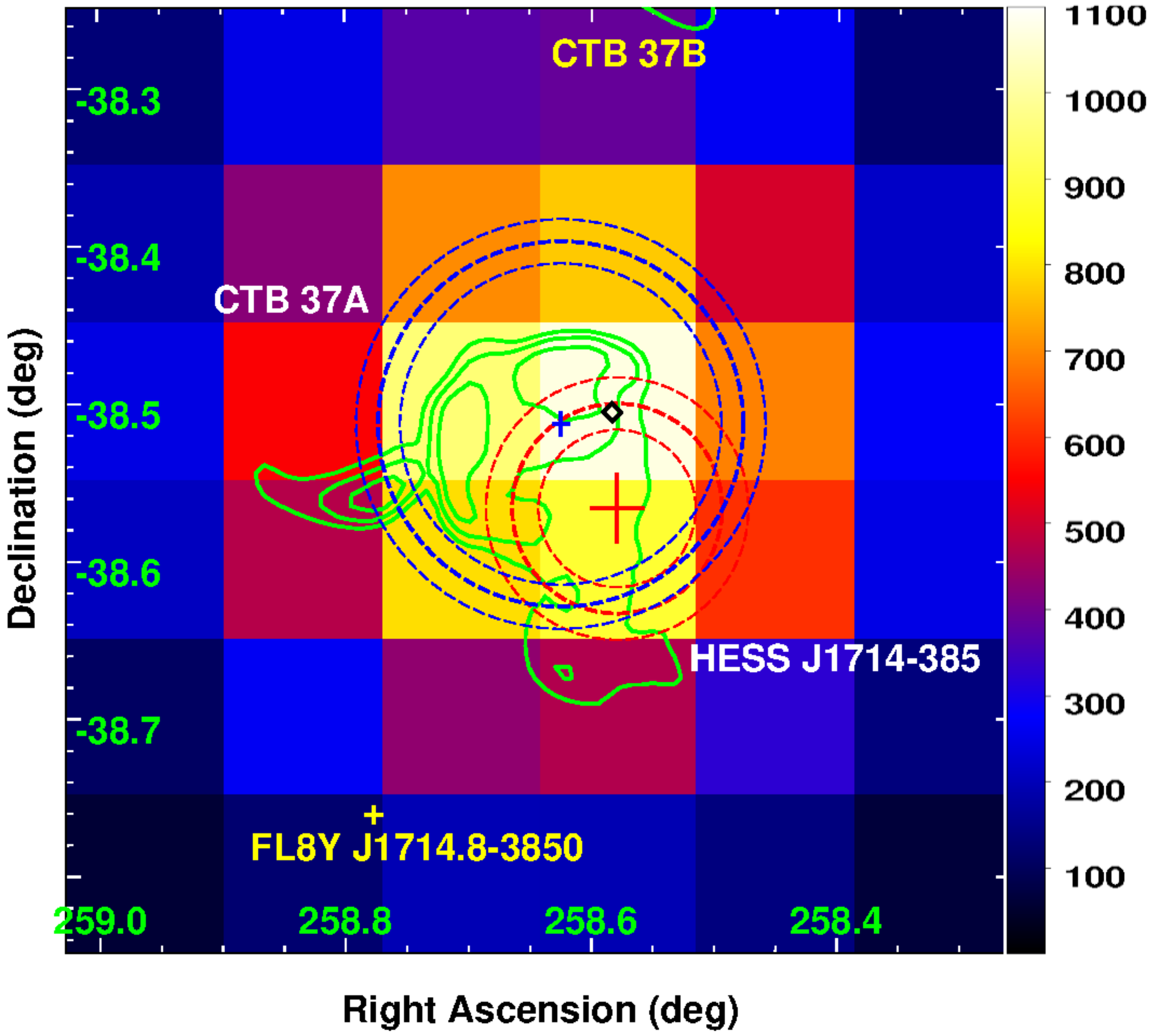}{0.41\textwidth}{}
          \fig{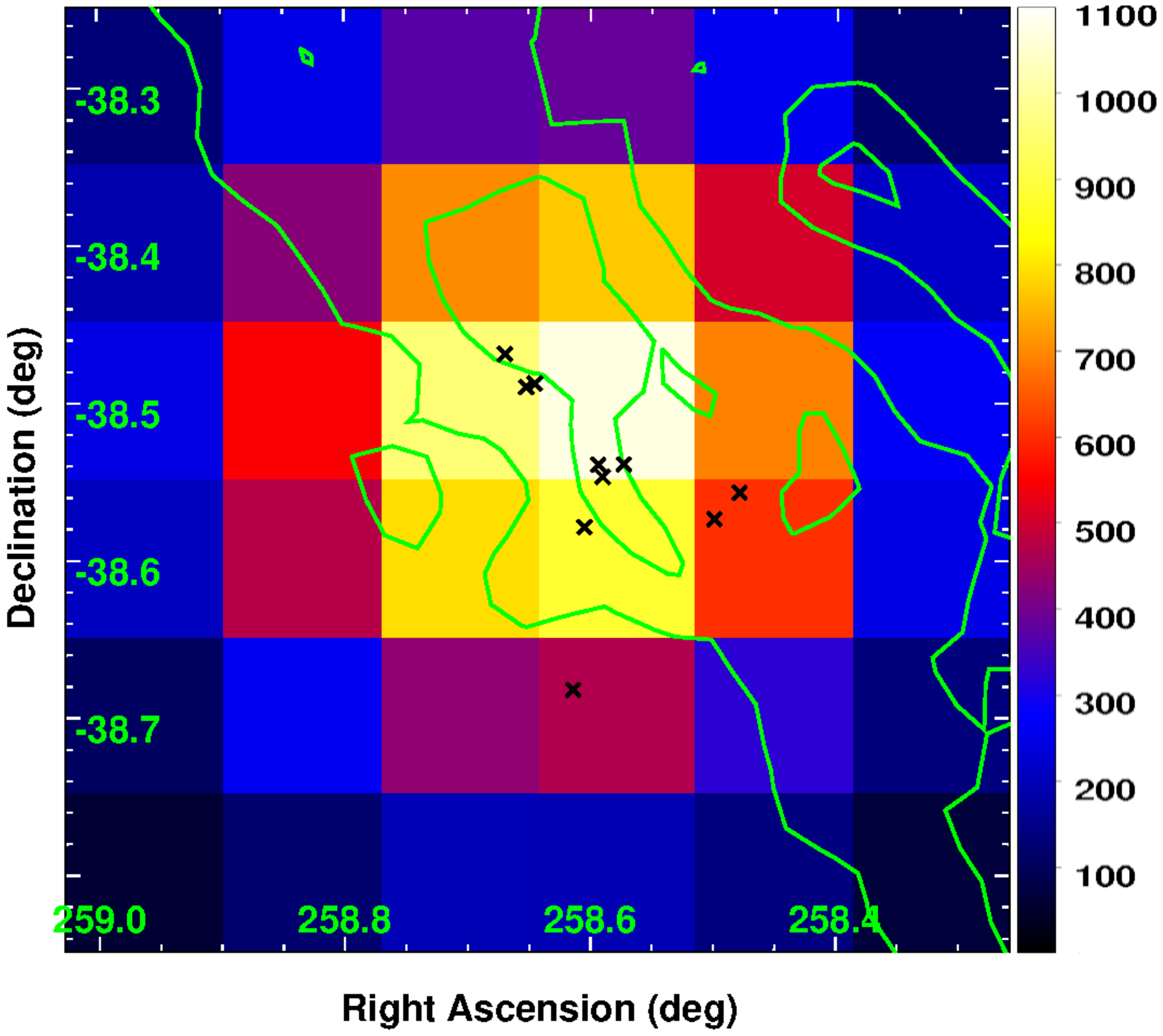}{0.417\textwidth}{}}                     
\caption{(Top left): $\it Fermi$-LAT counts maps in 0.2$-$200 GeV around the SNR CTB 37A in the entire ROI and using all data with a pixel size of $0\fdg05$ smoothed by a Gaussian kernel of $0\fdg05$.~(Top right): Same as the top left panel, but for the optimized data set (68\% PSF containment $<$ 1\fdg7, as defined in the text) and a zoomed view of the inner 5$^{\circ} \times 5^{\circ}$ region centered on the SNR position. (Bottom left): Test Statistic (TS) map in the vicinity of CTB 37A in the 1$-$200 GeV energy range. The 843 MHz radio contours from $\it MOST$ are overlaid in green.~The position of CTB 37A, HESS J1714$-$385, and their statistical errors, are marked with blue and red vertical crosses, respectively. The 68\% containment radii obtained by the Gaussian spatial model for CTB 37A and the HESS source~\citep{Aharonian08} are shown with thick dashed blue and red circles, respectively. The inner and outer radii in each case represent the statistical errors on the fitted extension. The black diamond indicates the position of the X-ray source CXOU J171419.8$-$383023. Yellow cross is the position of FL8Y J1714.8$-$3850. (Bottom right): \textit{Mopra} ${}^{12}$CO($J$=1$-$0) emission contours in green within the velocity range of $-$70 to~$-$50 km s$^{-1}$ are overlaid on the TS map (same as the bottom left panel).~The black crosses correspond to the position of the detected OH (1720 MHz) masers~\citep{Frail96}.\label{fig:Cmap-excessMap}}
\end{figure*}

After accounting for the systematic uncertainties associated with the localization due to the Galactic diffuse model and the LAT PSF (as $\sqrt{\sigma^{2}_{\rm IEM} + \sigma^{2}_{\rm PSF}}$$\,$), we then implemented two empirical corrections by multiplying the total uncertainty by a correction factor of 1.05 (as in 3FGL) and adding the absolute 95\% error of $0\fdg0075$ (as in 3FHL) in quadrature. The spatial properties of CTB 37A for the three tested models, along with their statistical and systematic errors, are summarized in Table~\ref{tab:Morphology_1}, in which the systematic effects dominate the total uncertainties. Table~\ref{tab:Morphology_2} lists the best-fit position and preliminary spectral properties of CTB 37A and the two closest sources, CTB 37B and FL8Y J1714.8$-$3850.~Our analysis confirms a 2D Gaussian morphology with a 68\% containment radius of 0\fdg116 and an extension significance of 5.75\,$\sigma$.~The obtained source size is smaller than the previously reported value by~\citet{Li17} by $\sim$40\%. This discrepancy can be explained by the contamination by the newly detected nearby source FL8Y J1714.8$-$3850 in our sky model. Our preliminary analysis based on the same data set but omitting FL8Y J1714.8$-$3850 also found a larger extension~\citep{Abdollahi17}. Figure \ref{fig:Cmap-excessMap} depicts the far and near vicinity of CTB 37A. The best-fit Gaussian extension (of 0\fdg116 in radius) is coincident with the bright part of the radio shell toward the east, and it is larger than the TeV emission of Gaussian width 0\fdg067 by $\sim$75$\%$.~Additionally, both X-ray and TeV sources are offset to the west from the geometric center of the remnant (bottom left panel of Figure \ref{fig:Cmap-excessMap}). The spatial properties of the remnant cannot preclude an association with a PWN. Further discussions in this regard will be given in Section~\ref{sec:Disc}.

\begin{deluxetable*}{lcccccccc}[!htbp]
\tablecaption{Best-Fit Spatial Properties of CTB 37A for Different Morphological Models Above 1 GeV \label{tab:Morphology_1}}
\tablecolumns{9}
\tablenum{1}
{
\tablehead{[-1.ex]
\colhead{Spatial Model} & \colhead{R.A.} & \colhead{Decl.} & \colhead{$r_{68}$} & \colhead{TS} & \colhead{TS$_{\mathrm{ext}}$} & \colhead{$N_{\rm dof}$}\\
[-2ex]
\colhead{} & \colhead{(deg)\,$\pm$\,stat\,$\pm$\,sys} & \colhead{(deg)\,$\pm$\,stat\,$\pm$\,sys} & \colhead{(deg)\,$\pm$\,stat\,$\pm$\,sys} & \colhead{} & \colhead{} & \colhead{}
}
\startdata
Point Source & 258.619\,$\pm$\,0.006\,$\pm$\,0.008 & $-$38.515\,$\pm$\,0.005\,$\pm$\,0.008 & ... & 1116 (1343) & ... & 5\\
Disk & 258.642\,$\pm$\,0.008\,$\pm$\,0.019 & $-$38.529\,$\pm$\,0.008\,$\pm$\,0.018 & 0.142\,$\pm$\,0.025\,$\pm$\,0.022 & 1511 (1902) & 19.61 (19.91) & 6\\
Gaussian & 258.625\,$\pm$\,0.007\,$\pm$\,0.017 & $-$38.513\,$\pm$\,0.008\,$\pm$\,0.011 & 0.116\,$\pm$\,0.014\,$\pm$\,0.017 & 1298 (1640) & 33.10 (42.48) & 6\\
\enddata}
\tablecomments{The best-fit positions for three tested models are given in J2000 epoch.~The 68\% containment radius for the disk and Gaussian models is defined as $r_{68}$ = 0.82$\it r$ and $r_{68}$ = 1.51$\it r$, respectively, where $\it r$ is the disk radius or the Gaussian $\sigma$.~The first and second errors on the spatial parameters correspond to the statistical and systematic, respectively.~The test statistic is evaluated from the likelihood ratio between two models with and without the source of interest, TS = 2~$\mathrm{log}$($\mathcal{L}$$_{\mathrm{on}}$/$\mathcal{L}$$_{\mathrm{off}}$).~A comparison of source extension for different hypotheses is provided by TS$_{\rm ext}$.~TS and TS$_{\rm ext}$ values for the analysis using the alternative diffuse model are given in parentheses.~$N_{\rm dof}$ corresponds to the number of degrees of freedom for each model.}
\end{deluxetable*}

\begin{deluxetable*}{lccccccc}[!htbp]
\tablecaption{Spatial and Spectral Properties of CTB 37A and Two Nearby Sources Using the Gaussian Fit \label{tab:Morphology_2}}
\tablecolumns{6}
\tablenum{2}
{
\tablehead{[-1.ex]
\colhead{Source Name} & \colhead{R.A.} & \colhead{Decl.} & \colhead{TS} & \colhead{Spectral Index ($\alpha$)} & \colhead{Curvature ($\beta$)}\\
[-2ex]
\colhead{} & \colhead{(deg)\,$\pm$\,stat} & \colhead{(deg)\,$\pm$\,stat} & \colhead{} & \colhead{} & \colhead{}
}
\startdata
CTB 37A & 258.625\,$\pm$\,0.007 & $-$38.513\,$\pm$\,0.008 & 1298 & 2.086\,$\pm$\,0.135 & 0.103\,$\pm$\,0.043\\
FL8Y J1714.8$-$3850 & 258.777\,$\pm$\,0.026 & $-$38.761\,$\pm$\,0.018 & 128 & 2.661\,$\pm$\,0.214 & 0.351\,$\pm$\,0.245\\
CTB 37B & 258.533\,$\pm$\,0.017 & $-$38.185\,$\pm$\,0.014 & 82 &1.611\,$\pm$\,0.127 & ...\\
\enddata}
\tablecomments{The last two columns give the spectral index $\alpha$ at the reference energy $E_{0}$ = 1 GeV and the curvature $\beta$ for a LogParabola spectrum. The errors are only statistical.}
\end{deluxetable*}

%%%%%%%%%%%%%%%%%%%%%%%%%%
\subsection{Spectral Analysis}\label{subsec:Spectrum}
After a global fit over the full energy range (0.2$-$200 GeV) using the best-fit model derived from the morphological analysis, the ROI model was reasonably optimized for generating the $\gamma$-ray spectrum of the SNR. A binned maximum likelihood analysis was performed in the full energy range, divided into nine logarithmically spaced energy bins by combining the four $\texttt{P8R2\char`_SOURCE\char`_V6}$ PSF event types in a joint likelihood approach. These four event types are based on the quality of the reconstructed direction, from the worst, PSF0, to the best, PSF3. Among different tested data sets, the data sample with a 68$\%$ PSF containment radius better than 1\fdg7 (hereafter optimized data set) strikes a fair balance between minimizing the Galactic diffuse emission contribution and maintaining sufficient photon statistics. The optimized data set corresponds to PSF0 events with $\it E \geq$ 1 GeV, PSF1 events with $\it E \geq$ 500 MeV, PSF2 events with $\it E \geq$ 316 MeV, and PSF3 events with $\it E \geq$ 200 MeV. The top right panel of Figure \ref{fig:Cmap-excessMap} depicts a zoomed view of the ROI surrounding the SNR for this optimized data set. A power law with the spectral index fixed at 2.0 is used to measure the spectral points, which makes the results independent of the spectral model in each energy bin.~As described above, only the normalizations of CTB 37A and the nearby bright sources with $>$\,9\,$\sigma$ significance (CTB 37B, RX J1713.7$-$3946, FL8Y J1714.8$-$3850, 3FGL J1718.1$-$3825, and 3FGL J1718.0$-$3726), as well as those of the Galactic and isotropic diffuse components, are adjusted.~Fixing the nearby sources results in underestimating the uncertainties due to the diffuse emission below 500 MeV.~The correction for energy dispersion is enabled for all model components except the Galactic and isotropic diffuse models, whose spectra are data-based. \par
\begin{figure}[ht!]
\plotone{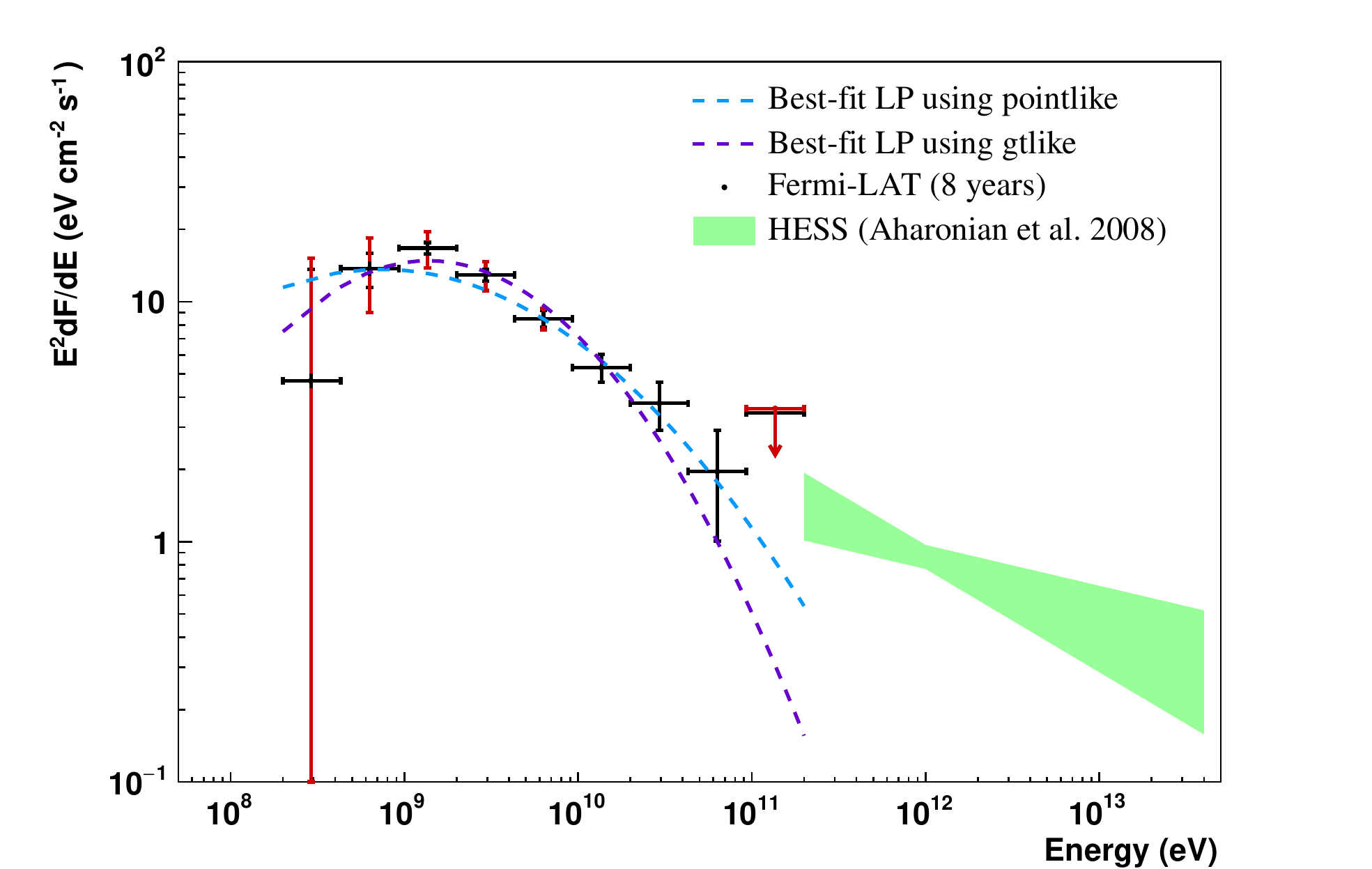}
\caption{Spectral energy distribution of CTB 37A is measured by $\textit{Fermi}$-LAT. Statistical and systematic uncertainties are given in black and red, respectively. An upper limit is calculated at 95\% confidence level using a Bayesian method in which the TS is less than 4. The systematic effects on the upper limit are taken from the larger of the errors caused by the Galactic diffuse emission and source spatial model. The blue curve shows the best fit with $\texttt{pointlike}$ above 1 GeV, extrapolated down to 0.2 GeV for better comparison. The purple curve indicates the global fit using $\texttt{gtlike}$ over the full energy range (0.2$-$200 GeV). The shaded region in green represents a power-law fit of the HESS J1714$-$385 measurements, taking into account the statistical errors only. \label{fig:SED}}
\end{figure}

 
Figure~\ref{fig:SED} shows the spectral energy distribution (SED) of CTB 37A. The statistical upper limit is calculated at 95\% confidence level using a Bayesian method~\citep[see, e.g.,][]{Helene83} when the detection is not significant (TS $<$ 4). Total systematic errors on the SED are included by adding in quadrature the uncertainties due to the Galactic diffuse model and the source spatial model as $\sqrt{\sigma^{2}_{\rm IEM} + \sigma^{2}_{\rm model}}$. The dominant uncertainties on the SED are the systematic errors due to the Galactic diffuse emission below 1 GeV and the statistical errors due to smaller statistics above 20 GeV, which together can explain the mismatch between two best fits by $\texttt{pointlike}$ and $\texttt{gtlike}$ (see Figure~\ref{fig:SED}). The total $\gamma$-ray energy flux of CTB 37A in the full energy range is calculated to be (1.08 $\pm$ 0.15$_{\rm stat}$ $\pm$ 0.19$_{\rm sys}$) $\times$ 10$^{-10}$ erg cm$^{-2}$ s$^{-1}$. The derived spectrum from the global fit with a spectral index of 1.92 $\pm$ 0.19 and a curvature of 0.18 $\pm$ 0.05 becomes softer above 10 GeV than the $\rm HESS$ spectrum with a spectral index of 2.3 $\pm$ 0.13~\citep{Aharonian08}, which suggests two or more particle populations at the origin of the GeV and TeV emission. The quoted errors on the spectral index are statistical only. \par
%%%%%%%%%%%%%%%%%%%%%%%%%%
\section{Discussion} \label{sec:Disc}
In the following discussion, based on the multiwavelength morphological and spectral characteristics of the remnant, we argue that CTB 37A should be classified as a composite SNR.~We then review our model, which is constructed to examine the reacceleration and compression of preexisting CRs in the adjacent molecular clouds as the primary source of the observed emission toward the SNR, following an analytical approach given by~\citet{Uchiyama10}. The $\mathtt{naima}$ package~\citep{Zabalza15} is used to perform the multiwavelength spectral fitting. We note that in all estimates presented below, a mean distance of 7.9 kpc to the SNR~\citep{Tian.Leahy12} is adopted. 
%%%%%%%%%%%%%%%%%%%%
\subsection{CTB 37A: A New Composite SNR} \label{subsec:CompositeSNR}
The LAT observation of the GeV $\gamma$-rays toward CTB 37A reveals a spatial extension approximately two times larger than that reported in the TeV range by~\citet{Aharonian08}. Furthermore, the LAT spectrum steepens above 10 GeV, so it is difficult to extend it smoothly into the harder TeV spectrum of HESS J1714$-$385. Taking these two facts together, it is very likely that CTB 37A is a composite system confining a nonthermal nebula inside its shell. Also, the VHE $\gamma$-ray source has an integrated $\gamma$-ray luminosity of $L_{\gamma(1-10~\rm TeV)} = 1.73 \times 10^{34}$ $\rm erg~s^{-1}$ in the energy range from 1 to 10 TeV for the mean distance of 7.9 kpc. It is in good agreement with the representative value $L_{\gamma(1-10~\rm TeV)} \approx 7 \times 10^{34}$ $\rm erg~s^{-1}$ for the observed middle-aged PWNe (between 7 and 23 kyr) as reported by~\citet{HESS18}. \par
Using $\it XMM$-$\it Newton$ and $Chandra$ observations of CXOU J171419.8$-$383023 of the northwestern rim of the SNR,~\citet{Aharonian08} derived a spectral index of $1.32^{+0.39}_{-0.35}$ for an extraction radius of $50^{\prime\prime}$.~\citet{Yamauchi14} analyzed the $\it Suzaku$ data and obtained a softer spectral index of $1.94^{+0.15}_{-0.14}$ for an extraction region of radius $\sim$$2^{\prime}$, quite typical index value for X-ray PWNe~\citep[see, e.g.,][]{Kargaltsev.Pavlov08}. The apparent discrepancy in the derived X-ray spectral index could be the result of the radiative losses of very energetic electrons with distance from the core of the emission. Not only fresh high-energy electrons but also accumulated aged cooled electrons can contribute to the $\it Suzaku$ spectrum. This spectral softening away from the core detected in CXOU J171419.8$-$383023 is additional evidence to support a composite classification for the SNR and suggests a PWN origin for the TeV emission~\citep[see, e.g.,][]{Slane00, Funk07}. Converting the X-ray flux to luminosity in the 0.5$-$10 keV band from $\it Suzaku$ data yields $L_{\rm X(0.5-10~\rm keV)} = 8.22 \times 10^{34}$ $\rm erg~s^{-1}$, which is compatible with the suggested $L_{\rm X(0.5-10~\rm keV)}\sim10^{35}~\rm erg~s^{-1}$ for the observed PWNe population~\citep{Gaensler.Slane06}. \par
In the PWN scenario, the larger size of the VHE emission compared with the X-ray one can be explained by the different lifetime of the emitting electrons. In this picture, high-energy electrons producing the synchrotron X-ray emission suffer severe radiative losses (so-called fast cooling). In contrast, low-energy electrons can survive longer and produce the TeV emission via IC scattering of ambient photons (slow cooling). The observed offset between the peak of TeV and X-ray emission ($\delta\theta\sim3.67^{\prime}$) could arise from an asymmetric reverse shock resulting from the expansion of the SNR into an initially inhomogeneous medium so that the reverse shock typically returns faster from the direction of higher ambient density~\citep[e.g.,][]{Blondin01}. \par
The offset of the nonthermal X-ray emission from the center of the radio SNR ($\delta\theta\sim2.0^{\prime}$) can be explained by the proper motion of the pulsar with a transverse velocity of $\simeq$ 750 $d_{7.9}$ $\tau_{6}^{-1}$ km s$^{-1}$, as supported by the tail-shape feature of the nonthermal X-ray emission~\citep{Aharonian08}. Here $\it d$$_{7.9}$ = $\it d$$_{\rm SNR}$/(7.9 kpc) and $\tau_{6}$ = $\tau_{\rm SNR}$/(6 kyr) are the normalized distance and age of the remnant, respectively. The assumed age of 6 kyr for the remnant is explained in Section~\ref{subsec:BroadbandModeling}. The measured pulsar's proper motion is faster than the sound speed in the shocked medium ($C_{\rm s}$ = 270 km s$^{-1}$), assuming an ideal gas with a mean temperature of $kT_{\rm X} = 0.64~\rm keV$ for the whole SNR from~\citet{Pannuti14}. \par 
All evidence (size, offset, index, and luminosity) is reasonably consistent with a composite system, suggesting that the TeV and hard X-rays arise from the putative associated PWN, whereas the observed radio, soft X-rays, and GeV $\gamma$-ray emission originates in the SNR shock. The consistency between the center of the SNR shell and the GeV peak provides additional justification for the proposed scenario. 
%%%%%%%%%%%%%%%%%%%%
\subsection{Modeling the Multiwavelength Emission from SNR CTB 37A} \label{subsec:BroadbandModeling}
Multiwavelength modeling of the remnant spectrum is conducted to probe the nature of the observed $\gamma$-ray emission toward the SNR. This approach includes the radio data at frequencies above 330 MHz taken from several surveys~\citep{Kassim91} and the LAT GeV spectrum from this work.\par
To explain the observed emission toward the system, we develop a numerical model in which acceleration of CRs by a strong shock propagating into the interstellar medium (ISM), as well as reacceleration of preexisting CRs in the nearby clouds by relatively weaker shocks, are considered. The model consists of three main parts: (i) production of energetic particles (including both protons and electrons), (ii) temporal evolution of the particle momentum distribution, and (iii) radiative mechanisms through the use of $\mathtt{naima}$. In the following two subsections, we provide the required physical parameters and discuss the details of our model.
%%%%%%%%%%%%%%%%%%%%%%%%%%
\subsubsection{Physical Parameters of the Model} \label{subsubsec:PhysicalParam}
To specify the required physical parameters for the model, we follow the approach described in~\citet{Devin18} and the references therein. Here, we assume that the remnant is in the Sedov (adiabatic) phase of evolution. The post-shock temperature is related to the temperature of the X-ray thermal gas as $kT_{\rm s} = kT_{\rm X}/1.27 = 0.50~\rm keV$~\citep{Borkowski01}. Moreover, the velocity of the blast wave assuming an adiabatic index of $\gamma = 5/3$ is given by:
\begin{equation}
v_{\rm s} = \Big(\frac{16\,kT_{\rm s}}{3\,\mu}\Big)^{1/2} \approx 650~\rm km~s^{-1},\label{eq1}
\end{equation}
where $\mu = 0.609\, m_{\rm H}$ is the mean atomic weight for a fully ionized gas. For the blast wave with a radius $R_{\rm s} = 10~\rm pc$, the age of the remnant can then be estimated using $t_{\rm s} = 2R_{\rm s}/5v_{\rm s} \approx 6000~\rm yr$. Consequently, the kinetic energy released through the SN explosion in terms of the shock dynamics is estimated by:
\begin{equation}
\frac{E_{\rm SN}}{10^{51}~\rm erg} = \Big(\frac{R_{\rm s}}{8.248~\rm pc}\Big)^{3} \Big(\frac{n_{0}}{1~\rm cm^{-3}}\Big)\,\Big(\frac{kT_{\rm s}}{1~\rm keV}\Big) = 0.57,\label{eq2}
\end{equation}
where $n_{0} = 0.64~\rm cm^{-3}$ is the ambient density.~All the input parameters (radius, density, and gas temperature) are taken from~\citet{Pannuti14}, assuming a distance of 7.9 kpc to the remnant as mentioned before. Equation~(\ref{eq2}) is derived from the standard Sedov solution $R_{\rm s}^{5} =$ $\xi\,E_{\rm SN}\, t_{\rm s}^{2}/\rho_{0}$, where $\xi = 2.026$ for the adiabatic index $\gamma = 5/3$, and $\rho_{0} = 1.4\,m_{\rm H}\,n_{0}$ is the ambient mass density. The uncertainty on the distance to the SNR translates directly into the uncertainties in the remnant age and explosion energy. \par 
The infrared observation of the SNR by~\citet{Andersen11} has revealed interactions with the surrounding clouds, in which radiative shocks occur, through the detection of [O\,{\footnotesize I}] emission at 63 $\mu$m, molecular and ionic lines. The ionic lines suggest moderately fast $\it J$-type shocks with velocities of $\sim$75$-$100 km s$^{-1}$ associated with compressed 10$^{3.5}$$-$10$^{4}$ cm$^{-3}$ media in the northern and southern shell of the remnant. Therefore, it is reasonable to consider the contribution of both the main shock expanding into the interstellar medium and the radiative shock driven into the clouds for the $\gamma$-ray production in our model. \par
A dynamic pressure equilibrium between the main shock and the radiative shock is expressed as: 
\begin{equation}
n_{\text{0,cl}} = k^{2}(v_{\text{s}}/v_{\text{s,cl}})^{2}\,n_{0}.\label{eq3}
\end{equation}
Thus for the main shock moving at velocity $v_{\rm s}$ = 650 km s$^{-1}$ into an ambient medium of density $n_{0} = 0.64~\rm cm^{-3}$ and a radiative shock with velocity $v_{\rm s,cl}$ = 100 km s$^{-1}$~\citep{Andersen11}, Equation~(\ref{eq3}) yields an upstream cloud density of $n_{\rm 0,cl}$ = 46 cm$^{-3}$, assuming a numerical factor $\it k$ = 1.3 as in~\citet{Uchiyama10}. \par
The downstream magnetic field strength in the cooled regions can be determined from
\begin{equation}
\frac{B_{\text{m}}^{2}}{8\pi} = k^{2}\rho_{0}v_{\text{s}}^{2},\label{eq4}
\end{equation}
balancing the magnetic pressure in the cooled regions with the ram pressure of the swept-up material in the shock~\citep{Hollenbach.McKee89}, in which 
\begin{equation}
B_{\text{m}} = \sqrt{2/3}\,(n_{\text{m}}/n_{\text{0,cl}})B_{\text{0,cl}}.\label{eq5}
\end{equation}
Thus, Equation~(\ref{eq4}) yields an amplified magnetic field strength of 520 $\mu$G in cooled regions of density $n_{\rm m} = 10^{3.5}$ cm$^{-3}$~\citep{Andersen11}. Such a strong magnetic field is consistent with the measurements of the OH (1720 MHz) maser features spatially coincident with the associated clouds in CTB 37A~\citep{Brogan00}. The $\sqrt{2/3}$ coefficient in Equation~(\ref{eq5}) refers to the tangential component of the magnetic field, which is amplified by compression and prevents the collapse of the radiative shock by providing pressure support. Using Equation~(\ref{eq5}), the upstream magnetic field strength in a cloud of density $n_{\rm 0,cl}$ is estimated to be $B_{\rm 0,cl}$ = 9.17 $\mu$G. \par
The upstream magnetic field strength and density in the clouds are related by
\begin{equation}
B_{\text{0,cl}} = b\,(n_{\text{0,cl}}/\text{cm}^{-3})^{1/2}~\mu\text{G},\label{eq6}
\end{equation}
where $\it b$ = $V_{\rm A}$/(1.84 km s$^{-1}$) is a dimensionless parameter related to the Alfv$\rm\acute{e}$n velocity $V_{\rm A}$.~As discussed in~\citet{Hollenbach.McKee89}, assuming that the velocity dispersion in molecular clouds is equal to the mean upstream Alfv$\rm\acute{e}$n velocity, $\it b$ ranges between about 0.3 and 3.~Here, the value of $\it b$ is constrained to $\sim$1.4 by Equation~(\ref{eq6}), which is consistent with the expectations.~Further, the downstream magnetic field (before radiative compression) can be obtained by $B_{\rm d,cl} = \sigma_{\rm B}\,B_{\rm 0,cl}$, where $\sigma_{\rm B} = \sqrt{(2\sigma^{2}+1)/3}$ is the magnetic compression ratio in terms of the shock compression ratio $\sigma$~\citep{Berezhko02}. It is assumed here that the magnetic field is fully isotropized. Then, the downstream magnetic field for a weakly modified shock ($\sigma$ = 4) is found to be $B_{\text{d,cl}}$ = 30 $\mu$G. Similarly, in the interstellar medium, the upstream magnetic field $B_{\rm ISM}$ = 1.1 $\mu$G and the downstream magnetic field $B_{\rm d,ISM}$ = 3.6 $\mu$G are derived for the ambient density of 0.64 cm$^{3}$. \par
Once the physical characteristics of the main shock, the radiative shock, and the media they are moving into are set, the spectral properties of emission from both components can be inferred following the same procedure as that adopted in~\citet{Parizot06}. In this approach, the acceleration timescale for $\sigma = 4$ is written as:
\begin{equation}
\tau_{\text{acc}} = 30.6\times k_{0}(E)\times E_{\text{TeV}}\,B_{2}^{-1}\,v_{\text{s,3}}^{-2}~~\text{yr},\label{eq7}
\end{equation}
which strongly depends on the shock velocity, and the synchrotron cooling timescale is estimated by
\begin{equation}
\tau_{\text{syn}} = (1.25\times 10^{3})\times E_{\text{TeV}}^{-1}\,B_{2}^{-2}~~\text{yr},\label{eq8}
\end{equation}   
where $B_{2}$ and $v_{\rm s,3}$ are the magnetic field and the shock velocity in units of 100 $\mu$G and 1000 km s$^{-1}$, respectively. In Equation~(\ref{eq7}), $k_{0}$ is the deviation of the diffusion coefficient from the Bohm value\footnote{The Bohm value is calculated when the particle mean free path is equal to its Larmor radius $r_{\rm L} = E/ZeB$, i.e., $D_{\rm Bohm} = r_{\rm L}c/3$.} ($D/D_{\rm Bohm}$), and it is found to be in the range from $\geq$1 to 10.~As the modeling of two young SNRs, Cas A~\citep{Zirakashvili14} and RX J1713.7$-$3946~\citep{Zirakashvili.Aharonian10}, shows, the particle acceleration in young SNRs proceeds close to the Bohm diffusion limit for which $k_{0}$ is about 1. However, in the case of evolved SNRs, the particle acceleration is not efficient, which means the $k_{0}$ parameter can be larger. In our model, $k_{0} = 10$ is used for the highest-energy electrons confined in the acceleration region.~The maximum achievable energy of particles in the main shock is calculated by matching the acceleration time either to the synchrotron cooling timescale or to the shock's age as $\tau_{\rm acc}$ = $\rm min$$\{\tau_{\rm syn}, t_{\rm s}\}$.~In the case of the radiative shock, as in~\citet{Uchiyama10}, the shock's age is the time elapsed since the clouds were shocked.~We approximate it by $t_{\rm c}$ = $t_{\rm s}/2$, which leads to $\tau_{\rm acc}$ = $\rm min$$\{\tau_{\rm syn}, t_{\rm c}\}$.~For the main shock moving into the interstellar medium of downstream magnetic field $B_{\rm d,ISM} = 3.6~\mu$G, we obtain a maximum energy $E_{\rm{max},\textit{e,p}}$ $\simeq$ 300 GeV. The slower radiative shock in the crushed clouds with downstream magnetic field of $B_{\rm d,cl}$ = 30 $\mu$G results in a lower value of $E_{\rm{max},\textit{e,p}}$ $\simeq$ 30 GeV.~Neither is limited by synchrotron cooling, so it is the same for electrons and protons. \par
Following~\citet{Uchiyama10}, we account for a break in the spectrum of accelerated particles in the radiative shocks, where the gas is partially neutral. This feature is associated with Alfv$\rm\acute{e}$nic turbulence and is due to particles accelerated along the upstream magnetic field in the clouds, which escape and result in a spectral steepening \citep{Malkov11}. \par
The additional adiabatic compression as $s = (n_{\rm m}/n_{\rm 0,cl})/\sigma$ in the radiative shocks boosts the momentum of accelerated particles by a factor $s^{1/3}$ close to 2.6. Table~\ref{tab:HadronicModelParam} summarizes the physical parameters in the model for both the main shock and the radiative shock in the clouds.~$E_{\rm max}$ and $E_{\rm b,\,Alf}$ are reported just downstream (before adiabatic compression).
%%%%%%%%%%%%%%%%%%%%%%%%%%
\subsubsection{Origin of the Gamma-Ray Emission: Accelerated or Reaccelerated CRs?}\label{subsubsec:OriginofCRs}
In the radiative shocks, the temporal evolution of the particle energy distribution is obtained by solving the transport equation, given by: 
\begin{equation}
    \frac{\partial N_{i}(E, t)}{\partial t}=-\frac{\partial}{\partial E} \big[\,\dot{E}\,N_{i}(E, t )\,\big]+Q_{i},\label{eq9}
  \end{equation}
where $N_{i}(E, t)$ is the density of particles of species $i$ (protons and electrons) per unit of energy, $\dot{E} \equiv dE/dt$ accounts for the total energy losses, and $Q_{i}$ denotes the injection rate of particles.~It is assumed in the model that particles are confined to the expanding shell.~Secondary $e^{\pm}$ generated through nuclear spallation are included. Contrary to the primaries, the injection rate of secondaries is time-dependent.~To solve the transport equation numerically, two quantities on the right-hand side of Equation~(\ref{eq9}), i.e., the energy losses and particle injection spectra, are specified as follows. \par
The energy loss mechanisms for electrons in the largely neutral clouds include synchrotron, bremsstrahlung, and ionization processes~\citep[][and references therein]{Prantzos11}.~In general, ionization is the dominant loss process at low energies, and the contribution of bremsstrahlung and synchrotron to energy loss becomes important in the intermediate and high energy ranges, respectively.~Protons in the radiative shocks suffer mainly from inelastic \textit{p}-\textit{p} interactions~\citep{Aharonian.Atoyan96} and ionization loss. \par 
We used the spectra of the Galactic CR protons $n_{\rm GCR,\it p}(p)$ and electrons+positrons $n_{\rm GCR,\it e}(p)$ from~\citet{Uchiyama10} as the initial seed particles for considering the reacceleration process by the shocks in the model.~The spectra of particles have been extended downward in energy to 20 MeV. It is much below the threshold energy of pion production, but it is necessary for computing their contribution to the total energy budget of the system. We assume that the CR properties in the clouds are the same as in the ISM. \par
The accelerated/reaccelerated CR spectrum $n_{\rm acc}(p)$ \citep[see Equation~(3) in][]{Uchiyama10} through the DSA process~\citep{Blandford.Eichler87} undergoes several modifications.
An exponential cutoff in momentum as a factor of $\mathrm{exp}[-(p/p_{\rm max})]$ is introduced to take into account the maximum attainable energy derived using Equations~(\ref{eq7}) and~(\ref{eq8}). The DSA proton spectrum steepens by one power above the break momentum $p_{\rm b,\,Alf}$ due to the evanescence of Alfv$\rm\acute{e}$n waves in the radiative shocks. Moreover, the adiabatic compression inside cooling regions behind the radiative shock results in an enhanced energy density as $n_{\rm ad}(p)=s^{2/3}\,n_{\rm acc}(s^{-1/3}\,p)$. \par
Using the particle density in the radiative shocks, the injection rate of particles into the transport process is determined by:
  \begin{equation}
   Q_{p,e}^{\rm r}(p)=\frac{f\,V_{\rm SNR}\,n_{\rm 0,cl}}{n_{\rm m}\,t_{\rm c}}\,n_{\rm ad}^{\rm r}(p),\label{eq10}
  \end{equation}
where the quantity $f$ is the filling factor of the clouds before they were crushed relative to the SNR volume $V_{\rm SNR}$. $t_{\rm c}$ represents the time spent by energetic particles in the shocked clouds.~As in~\citet{Uchiyama10}, it is assumed that the shock velocity remains constant inside the clouds. \par
In addition to the primary particles, secondary $e^{\pm}$ are also produced in the clouds of density $n_{\rm m}$ through hadronic interactions and, consequently, production and decay of charged pions.~Solving Equation~(\ref{eq9}) numerically, the proton spectrum in the crushed clouds $N_{p}^{\rm r}(p,t)$ serves as the source term for calculating the spectrum of secondary $e^{\pm}$. The injection rate of secondaries is computed as
\begin{equation}
   Q_{e^{\pm}}^{\rm r}(E,t)=n_{\rm m}\,c\,\epsilon_{\rm M}\,\int\!dT_{p}\,N_{p}^{\rm r}(T_{p},t)\,\frac{d\sigma(E,T_{p})}{dE},\label{eq11}
  \end{equation}
where $c$ is the speed of light, and $\epsilon_{M}$ is the nuclear enhancement factor to take into account the effects of heavier nuclei ($\it A$ $>$ 1) in both CRs and the target material~\citep{Mori09}.~The last term $d\sigma(E,T_{p})/dE$ is the energy-dependent differential cross section of $e^{\pm}$ generation from incident protons of kinetic energy $T_{p}$ as parametrized by~\citet{Kamae06}.~Contrary to the primaries, the injection rate of secondary $e^{\pm}$ is time-dependent.~A low-energy cutoff at a kinetic energy of 1 MeV is assumed for secondaries. \par
Using all the required ingredients, the numerical solution of the transport equation provides the final spectrum of accelerated CR species $N(p,t)$.~Equation~(\ref{eq9}) is solved at $t_{\rm c}$ for both preexisting particles in the clouds and the secondaries. We then apply $\mathtt{naima}$ for computing nonthermal radiation processes and fitting the multiwavelength data toward the remnant.~The radiation from neutral pion-decay, bremsstrahlung, synchrotron, and IC are included. The interstellar radiation field (ISRF) at the location of the remnant is taken from GALPROP~\citep{Porter17} as seed photons for IC. Energy density and temperature of the ISRF in the near-infrared band, far-infrared band, and the cosmic microwave background (CMB) photons are ($U_{\rm NIR}$ = 1.84 eV cm$^{-3}$, $T_{\rm NIR}$ = 3484.16 K), ($U_{\rm FIR}$ = 0.81 eV cm$^{-3}$, $T_{\rm FIR}$ = 28.98 K), ($U_{\rm CMB}$ = 0.26 eV cm$^{-3}$, $T_{\rm CMB}$ = 2.72 K), respectively. The $\gamma$-rays from \textit{p}-\textit{p} interactions are rescaled by the nuclear enhancement factor $\epsilon_{M}$. \par
Additionally, we considered the contribution of acceleration of fresh CRs at the blast wave.~At this site, the energy losses of particles are negligible, and there are no secondaries, so we use the analytic approximation of the solution of the transport equation.~In our model, the energy distribution of electrons is approximated by a power law with a break associated with the aging of the particles and an exponential cutoff as: 
\begin{equation}
    N_{e}(E)\,=A_{e}\,\text{exp}{(-E/E_{\rm max})}
    \begin{cases}
        {(E/E_{0})}^{-\Gamma_{e,1}} & E \leq E_{\rm b,\,syn}\\
       {(E_{\rm b,\,syn}/E_{0})}^{\Gamma_{e,2}-\Gamma_{e,1}}\,{(E/E_{0})}^{-\Gamma_{e,2}} & E > E_{\rm b,\,syn},\label{eq12}
    \end{cases}
  \end{equation}
in which the electron spectral index changes by unity ($\Gamma_{e,2} - \Gamma_{e,1} = 1$) after the break energy $E_{\rm b,\,syn}$. The cutoff energy $E_{\rm max}$ corresponds to the maximum achievable particle energy within the acceleration process, and the reference energy $E_{0}$ is set to 1 GeV. For the proton spectrum, a power law with spectral index $\Gamma_{p}$ and an exponential cutoff at $E_{\rm max}$ in the form
\begin{equation}
    N_{p}(E)=A_{p}\,{(E/E_{0})}^{-\Gamma_{p}}\,\text{exp}{(-E/E_{\rm max})} \label{eq13}
  \end{equation}
is assumed.~The proton and electron spectral indices $\Gamma_{p}$\,=\,$\Gamma_{e,1}$\,=\,2 are set as predicted by the DSA mechanism.~Equations~(\ref{eq12}) and~(\ref{eq13}) are in terms of energy, as defined in $\mathtt{naima}$.~The $\gamma$-rays from \textit{p}-\textit{p} interactions are enhanced by the factor $\epsilon_{M}$.~$E_{\rm b,\,syn}$ is set by equating the synchrotron cooling timescale with the shock's age through $\tau_{\rm syn} = t_{\rm s}$ in the main shock.~Due to the weak downstream magnetic field at the main shock, the synchrotron cooling is insignificant, and the resulting spectral break is beyond the cutoff energy ($E_{\rm b,\,syn}\propto B_{\rm d,ISM}^{-2}$). So, the electron distribution is a single power law as Equation~(\ref{eq13}).~The total kinetic energy of the accelerated protons is set to 10\%\,($1 - f$)\,$E_{\rm SN}$, i.e., $W_{p}$ = 1.58$\times10^{49}$ erg. The energy budget of accelerated electrons is set to 5\% that of the protons. \par
\begin{figure*}[!htbp]
\plotone{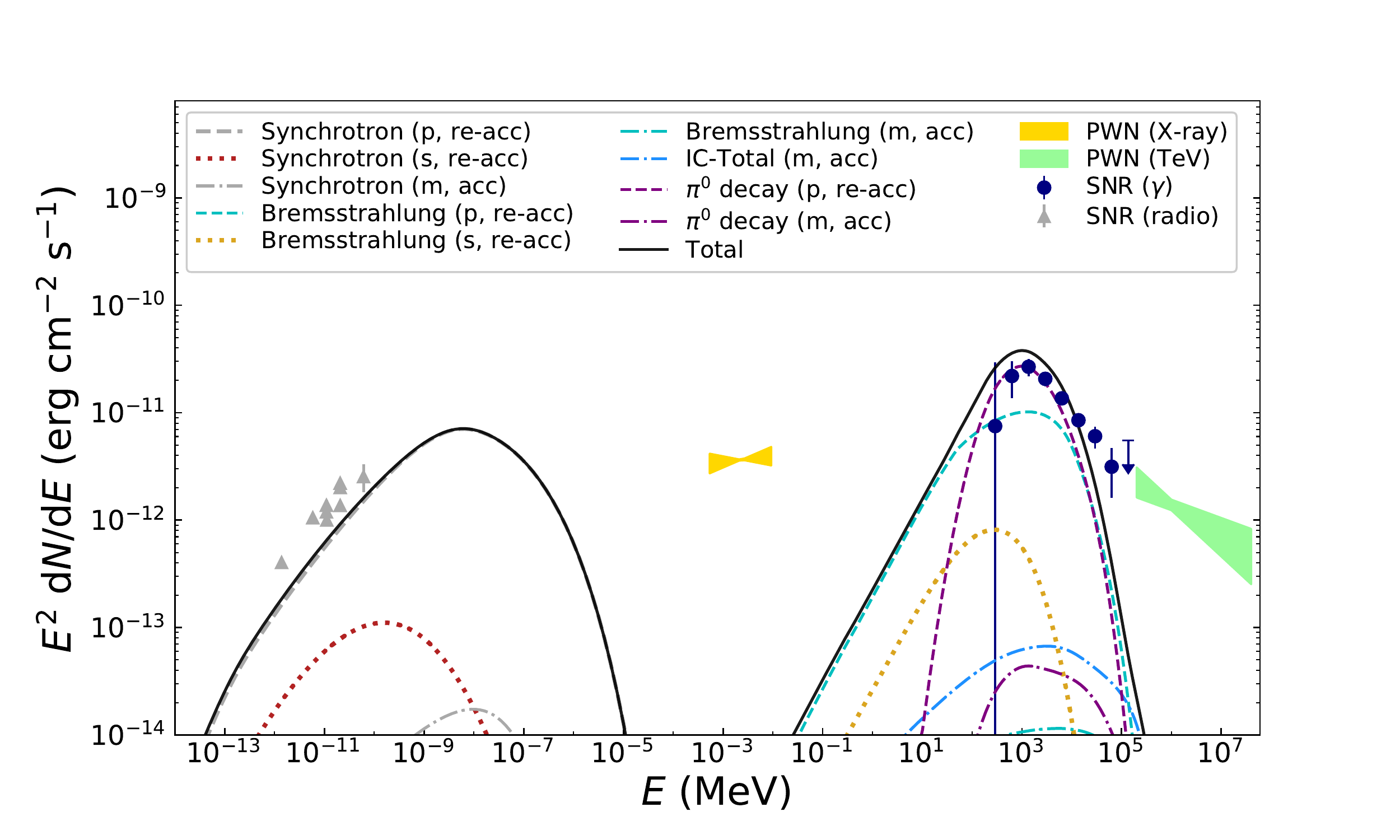}
\caption{Acceleration/reacceleration scenario for the multiwavelength modeling of the emission toward CTB 37A, in which dashed lines correspond to the emissions from primaries (p) and dotted lines correspond to those of secondaries (s) in the radiative shocks.~Dash-dotted lines represent the emissions from fresh CRs accelerated at the blast wave. The total model from the summation of all emissions is shown in black.~The IC emission from the reaccelerated electrons in the clouds is insignificant, with a peak at $\sim$\,2$\times$10$^{-15}$ erg~cm$^{-2}$~s$^{-1}$.~The radio data points are a combination of several observations at frequencies above 330 MHz, taken from Table 2 of~\citet{Kassim91}. The $\it Suzaku$ X-ray and \rm H.E.S.S. TeV spectra are from~\citet{Yamauchi14} and~\citet{Aharonian08}, respectively. The error bars on the LAT measurement are calculated by adding statistical and systematic errors in quadrature. \label{fig:HadronicDominated}}
\end{figure*}

Because all physical parameters (except the energy budget of CRs through the acceleration process) are constrained by observations, the level of gamma-ray emission at the radiative shocks is defined only by the filling factor $f$ of the clouds and is fit to the data.~The resulting factor of 0.72 is large but not impossible.
The fit requires 0.36\% of $E_{\rm SN}$ to be transferred to the radiative shocks. In this model, the maximum energy $E_{\rm max}$ with the assumed $k_{0} = 10$ provides a reasonable fit to the data. The resulting spectrum using these values implies that the radiative shells driven into the dense clouds are the dominant contributor to the observed radio and GeV emission (see dashed-line curves in Figure~\ref{fig:HadronicDominated}), and the emission from the blast wave is negligible, similar to SNR W44 as tested by our model. Secondary particles are found to contribute insignificantly to the total radio and GeV emission.~Moreover, the model predicts that 13\% of the energy injected into the protons was lost since the clouds were shocked. Because the physical parameters are entirely constrained in our model, it is impossible to fit the ratio of the radio to gamma-ray data exactly. The model ratio on Figure \ref{fig:HadronicDominated} is about twice too low. Having a larger magnetic field in the radiative shocks (i.e., a larger pressure in the SNR) would improve that. We note that for the results presented here, the distance to the remnant is the main source of uncertainty in our calculations.

\begin{deluxetable*}{lccccccc}[!htbp]
\tablecaption{Parameters for the Model Shown in Figure \ref{fig:HadronicDominated} \label{tab:ReAcceleration}}
\tablecolumns{8}
\tablenum{3}
\tablewidth{0pt}
\tablehead{
\colhead{Emission Region} & \colhead{$\it n$ (cm$^{-3}$)} & \colhead{$B$ ($\mu$G)} & \colhead{$E_{\rm max}$ (GeV)} & \colhead{$E_{\rm b,\,Alf}$ (GeV)} & \colhead{$f$} & \colhead{$W_{p}$ ($\times10^{48}$ erg)} & \colhead{$W_{e}$ ($\times10^{47}$ erg)}}
\startdata
Main Shock & 0.64 & 3.6 & 300 & \rm{no value} & 0.3 & 0.6 (15.8) & 0.3 (7.9) \\
Radiative Shock & 46 & 30 &  &  &  &  &  \\
Cooled Regions & 3160 & 520 & 30 & 11$^{\dagger}$ & 0.7$^{*}$ & 1.9 & 1.4 \\
\enddata
\tablecomments{Model parameters for both the main shock ($v_{\rm s} = 650$ km s$^{-1}$) and radiative shocks ($v_{\rm s,cl} = 100$ km s$^{-1}$) in SNR CTB 37A. The ambient medium properties (density and magnetic field) are taken from the observational data. The cutoff energy $E_{\rm max}$ and the break energy $E_{\rm b,Alf}$ are derived as in~\citet{Parizot06} and~\citet{Malkov11}, respectively. $W_{p}$ and $W_{e}$ are the energy input to reaccelerated protons and (primary) electrons, respectively.~Values in parentheses are the energy budget of CRs through the acceleration of fresh particles in the blast wave. A parameter marked with an asterisk is fit to the data. The spectral break marked with a dagger is caused by the Alfv$\rm\acute{e}$n wave evanescence. \label{tab:HadronicModelParam}}
\end{deluxetable*}

%%%%%%%%%%%%%%%%%%%%%%%%%%
\section{Comparison with other middle-aged SNRs} \label{sec:Comparison}
As reported in Table 5 of~\citet{Acero2016}, among all surveyed Galactic SNRs by $\it Fermi$-LAT, only 11 of them have associated OH masers, which are signposts of SNR$\slash$MC interaction.~The measured magnetic field strength of $\sim$ mG toward those masers is consistent with those for shock-compressed clouds~\citep[see, e.g.,][]{Brogan00}.~Table \ref{tab:ListofMidAgeSNRs} lists three interacting SNRs along with their physical parameters, for which the reacceleration of particles, followed by compression in radiative shocks ($v_{\rm s,cl}\geq100$ km s$^{-1}$), is believed to be at the origin of hadronic $\gamma$-ray emission.~The compressed gas and amplified magnetic field by the fast shocks in the interaction region ($v_{\rm s,cl}\gtrsim100$ km s$^{-1}$) play a crucial role in the hadronic nature of their $\gamma$-ray emission. \par
As shown in Table \ref{tab:ListofMidAgeSNRs}, the maximum achievable energy in CTB 37A is lower than the two other SNRs. It is mainly due to a shorter acceleration time compared with the two other SNRs, a lower downstream magnetic field in its associated clouds compared with that in W44, and a slower shock in the clouds relative to MSH 15$-$56 ($E_{\rm max}\propto t_{\rm c}\,v_{\rm s,cl}^{2}\,B_{\rm d,cl}$). \par
In CTB 37A, the emission from secondaries does not exceed the primaries' spectra, contrary to the results in~\citet{Uchiyama10} for W44.~Interestingly, as in~\citet{Lee15}, our model results in a subdominant contribution of secondaries in SNR W44 using the same physical parameters as listed in Table \ref{tab:ListofMidAgeSNRs}.~\citet{Lee15} have attributed the observed difference to the fact that a time-dependent shock velocity is assumed in their model. However, we have followed the same approach as~\citet{Uchiyama10}, in which the dynamics of the remnant are not taken into account.~The difference in the flux of secondary electrons may stem from how the nonthermal cooling in the radiative shock is computed in~\citet{Uchiyama10}.

\begin{deluxetable*}{lccc}[!htbp]
\tablecaption{Physical Properties of Three Middle-aged SNRs Explained with the Crushed Clouds Scenario \label{tab:ListofMidAgeSNRs}}
\tablecolumns{4}
\tablenum{4}
\tablewidth{0pt}
\tablehead{
 \colhead{Physical Parameters} & \colhead{CTB 37A} & \colhead{MSH 15$-$56} & \colhead{W44} \\
\colhead{} & \colhead{} & \colhead{} & \colhead{} 
}
\startdata
SNR Dynamics &  &  & \\
\hline
$\tau$ (kyr) & 6 & 16.5 &10 \\
$\it d$ (kpc) & 7.9 & 4.1 & 2.9 \\
$\it R$ (pc) & 10 & 21 &12.5 \\
$E_{\rm SN}$ ($\times$$10^{51}$ erg) & 0.57 & 0.5 & 5 \\[0.1cm]
\hline
Gas Properties in Clouds &  &  & \\
\hline
$v_{\rm s,cl}$ (km s$^{-1}$) & 100 & 150 & 100 \\
$\it n_{\rm 0,\rm cl}$ ($\rm cm^{-3}$) & 46 (183) & 2 (8) & 200 (800) \\ 
$\it B_{\rm 0,\rm cl}$ ($\mu\rm G$) & 9 (30) & 4 (14) & 25 (83) \\
$\it n_{\rm m}$ ($\rm cm^{-3}$) & 3160 & 90 & 10620 \\ 
$\it B_{\rm m}$ ($\mu\rm G$) & 520 & 160 & 1080 \\[0.1cm]
\hline
Gas Properties in Blast Wave &  &  & \\
\hline
$v_{\rm s}$ (km s$^{-1}$) & 650 & 500 & 490 \\
$\it n_{\rm 0}$ ($\rm cm^{-3}$) & 0.64 (2.56) & 0.1 (0.4) & 5 (20) \\ 
$\it B_{\rm ISM}$ ($\mu\rm G$) & 1.1 (3.6) & 3.0 (10.0) & 4.0 (13.1) \\[0.1cm]
\hline
Spectral Parameters in Radiative Shocks &  &  & \\
\hline
$p_{\rm c}$ ($\rm GeV$/c) & 30 & 82.7 & 122 \\
$p_{\rm b}$ ($\rm GeV$/c) & 11.3 & 15.2 & 7 \\
$\alpha_{\rm r}$ & 0.50 & 0.34 & 0.37 \\
$^{\dagger}$References & (1) & (2) & (3) \\
\enddata
\tablenotetext{\dagger}{The numbers refer to the following references: (1) This work, (2) \citet{Devin18}, (3) \citet{Uchiyama10}}
\tablecomments{\,$\tau$ is the age, and $\it d$ is the distance to SNR.~Upstream density and magnetic field in the clouds are given by $n_{0,\rm cl}$ and $B_{0,\rm cl}$, respectively. These parameters in the ISM are presented by $n_{0}$ and $B_{\rm ISM}$, respectively. The numbers in the parentheses correspond to those in the downstream regions. The two parameters $n_{\rm m}$ and $B_{\rm m}$ correspond to those in the cooled regions.~$\alpha_{\rm r}$ refers to the radio spectral index.~The spectral break $p_{\rm b}$ is due to Alfv$\rm\acute{e}$n wave damping in CTB 37A and W44, and to synchrotron cooling in MSH 15$-$56.~All other parameters are the same as those explained in the text. 
}
\end{deluxetable*}
%%%%%%%%%%%%%%%%%%%%%%%%%%
\section{Conclusions} \label{sec:Conclusions}
Using 8 yr of Pass 8 $\it Fermi$-LAT data, we have studied the nature of the $\gamma$-ray emission in the direction of the CTB 37A system. The morphological analysis of the source using all data above 1 GeV revealed an extended emission that is best modeled by a Gaussian distribution with $r_{68}$\,=\,0$\fdg$116$\,\pm\,$0$\fdg$014$_{\rm stat}$$\,\pm\,$0$\fdg$017$_{\rm sys}$ at a significance of 5.75\,$\sigma$. The measured angular extension is comparable with the radio size, while it is larger than the extension of the TeV emission from HESS J1714$-$385 by $\sim$75$\%$. The GeV emission is centered on the radio SNR and offset from both the nonthermal X-ray and the TeV emission. The spectral analysis of the remnant using an optimized data set with $r_{68}$ less than 1\fdg7 over the full energy range (0.2$-$200 GeV) showed that the GeV spectrum steepens above 10 GeV compared with the $\rm HESS$ spectrum, which strengthens two different origins for the GeV and TeV emission toward this system. The detected pulsar inside the system confirms a composite class of the SNR, as we proposed.\par
Assuming the SNR is in the Sedov stage, we examined a scenario in which both the acceleration of Galactic CRs and the reacceleration of preexisting CRs are considered. The maximum energy of particles through the DSA process reaches 300 GeV in the blast wave and 30 GeV in the radiative shocks. The low-density ambient medium leads to faint emission from the blast wave.~On the contrary, the dense clouds interacting with the SNR can explain well both the radio and GeV spectra through the reacceleration of CRs followed by radiative compression. For all physical parameters fixed at their values from the observational data, a reasonable fit to the radio and GeV spectra can be obtained if the clouds occupied most of the SNR volume prior to explosion. The energy left in the CR protons implies that 13\% of the injected energy was lost since the clouds were shocked. Moreover, the contribution of secondaries is subdominant compared with the primaries. \par
The accumulated \textit{Fermi}-LAT data will provide more candidates in which the reacceleration of CRs is the dominant process. The Cherenkov Telescope Array will resolve the CTB 37A system and clarify the relation between the SNR and the PWN.
\acknowledgments
$\it Acknowledgements.$ The \textit{Fermi} LAT Collaboration acknowledges generous ongoing support
from a number of agencies and institutes that have supported both the
development and the operation of the LAT as well as scientific data analysis.
These include the National Aeronautics and Space Administration and the
Department of Energy in the United States, the Commissariat \`a l'Energie Atomique
and the Centre National de la Recherche Scientifique / Institut National de Physique
Nucl\'eaire et de Physique des Particules in France, the Agenzia Spaziale Italiana
and the Istituto Nazionale di Fisica Nucleare in Italy, the Ministry of Education,
Culture, Sports, Science and Technology (MEXT), High Energy Accelerator Research
Organization (KEK) and Japan Aerospace Exploration Agency (JAXA) in Japan, and
the K.~A.~Wallenberg Foundation, the Swedish Research Council and the
Swedish National Space Board in Sweden.
 
Additional support for science analysis during the operations phase is gratefully
acknowledged from the Istituto Nazionale di Astrofisica in Italy and the Centre
National d'\'Etudes Spatiales in France. This work performed in part under DOE
Contract DE-AC02-76SF00515.
\facility{\textit{Fermi}-LAT.}
\software{\textit{Fermi} Science Tools\footnote{\url{https://fermi.gsfc.nasa.gov/ssc/data/analysis/software/}} (version~v11r05p02), $\mathtt{naima}$ package~\citep{Zabalza15}}.



\end{document}